\documentclass{czjphys}         % for LaTeX 2e

\usepackage{graphicx} % standard LaTeX graphics for including eps-figure files
\usepackage{cite}     % this package collapses a list of three or more consecutive reference numbers into a range
\usepackage{amsfonts}

\def \d {{\rm d}}
\def \e {e}

\begin{document}
\title{Evolution of high-frequency gravitational waves in some cosmological models}  
\authori{Otakar Sv\'{\i}tek and Ji\v{r}\'{\i} Podolsk\'{y}}      \addressi{Institute of Theoretical Physics, Charles University in Prague,\\
Faculty of Mathematics and Physics, V~Hole\v{s}ovi\v{c}k\'ach 2,\\ 
180~00 Praha 8, Czech Republic}
\authorii{}     \addressii{}
\authoriii{}    \addressiii{}
\authoriv{}     \addressiv{}
\authorv{}      \addressv{}
\authorvi{}     \addressvi{}
%
%Page headings:
\headauthor{Otakar Sv\'{\i}tek and Ji\v{r}\'{\i} Podolsk\'{y}}            % page heading on the even pages
\headtitle{Evolution of high-frequency gravitational waves in some cosmological models}             % page heading on the odd pages
\lastevenhead{Otakar Sv\'{\i}tek and Ji\v{r}\'{\i} Podolsk\'{y} et al.: Evolution of high-frequency gravitational waves \ldots} % p. h. on the last page if even
     % max. 2 codes of Physics and Astronomy 
               % Classification Scheme
 % lowercase letters

%%%%%%%%%%%%%% FOR EDITORIAL USE ONLY!!! %%%%%%%%%%%%%%%
\refnum{}%\total{}\type{}
\daterec{}    %;\\ final version }
\issuenumber{}  \year{2006}
\setcounter{page}{1}
%\firstpage{1}
%\lastpage{000}
%\makefirsttitle
%%%%%%%%%%%%%%%%%%%%%%%%%%%%%%%%%%%%%%%%%%%%%%%%%%%%%%%%
\maketitle

\begin{abstract}
We investigate Isaacson's high-frequency gravitational waves which propagate in some relevant cosmological models, in particular the FRW spacetimes. Their time evolution in Fourier space is explicitly obtained for various metric forms of (anti--)de~Sitter universe. Behaviour of high-frequency waves in  the anisotropic Kasner spacetime is also described.
\end{abstract}

\section{Introduction}
During the last years a growing attention has been paid to analysis of gravita-tional-wave perturbations propagating on cosmological backgrounds. The motivation stems not only from theoretical considerations concerning the influence of primordial gravitational waves on the cosmological evolution, in particular on the CMB anisotropies,
but also from the emerging field of gravitational-wave astronomy which stimulates studies on 
stochastic backgrounds of gravitational waves (see, e.g., \cite{bardeen,ellisbruni,oshea,katz,lidsey,melchiorri,mollerach,tomita,maggiore,zhang} for number of references).

Many years ago Isaacson introduced and developed an interesting gauge invariant perturbation approach that enables one to study properties of high-frequency gravitational waves, together with their influence on the background in which they propagate \cite{isaac,maccallumtaub}. The complete Isaacson's formalism thus consistently combines the equation for  high-frequency waves with the equation describing back-reaction of the waves on the background. Recently we investigated this effect using the WKB form of the gravitational-wave perturbations \cite{podolsky-svitek} and we presented relations between these solutions and some exact radiative cosmological models. 

It is  rather surprising that the Isaacson wave equation itself did not attract much attention. To our knowledge it was solved (in fact, by several authors \cite{isaac,choquet2,podolsky-svitek,maccallumtaub,taub2,hogan}) 
basically in case of the Robinson-Trautman spacetimes, the Vaidya solution in particular. However, such spacetimes are usually considered as a model of spherical gravitational waves generated by a ``local source'' rather than a  cosmological model. 
The purpose of our present work is to solve the Isaacson wave equation describing propagation of high-frequency waves on  ``genuine'' cosmological backgrounds (without, however, considering the back-reaction). 

After summarizing the Isaacson formalism (section~2) we apply it to some fundamental cosmological models (section~3). General family of the Friedmann--Robertson--Walker spacetimes is studied in subsection~3.1. By decomposing the high-frequency perturbations into tensor harmonics we derive the evolution equation for various permitted gravitational waves. In particular, illustrative explicit solutions are presented for the de~Sitter  and anti--de~Sitter universes using various metric forms of these spacetimes. In subsection~3.2 we investigate possible high-frequency gravitational waves in the Kasner universe which is a classic representative of anisotropic but homogeneous cosmological models.

%\newpage

\section{The Isaacson formalism}
Isaacson's perturbation method \cite{isaac} is based on decomposition of the spacetime metric $g_{\mu\nu}$ into the background metric
$\gamma_{\mu\nu}$  and its perturbation $h_{\mu\nu},$
\begin{equation}\label{x1.1.1}
g_{\mu\nu}=\gamma_{\mu\nu}+h_{\mu\nu}\ ,
\end{equation}
where, in a suitable coordinate system, ${\gamma_{\mu\nu}=O(1)}$ and
${h_{\mu\nu}=O(\epsilon)}$. By definition, $f=O(\epsilon^n)$ if there exists a constant ${C>0}$ such that
${|f|<C\epsilon^n}$ as ${\epsilon\to 0}$. 
The assumption ${h_{\mu\nu}=O(\epsilon)}$ thus does \emph{not} automatically imply that 
${h_{\mu\nu}\sim\epsilon}$. The spectrum of possible waves is not a priori restricted, 
it is only required that their amplitudes  fall to zero at least linearly with~$\epsilon$, 
i.e. ${|h_{\mu\nu}(\epsilon)|<C\epsilon}$.

Physically, the non-negative dimensionless parameter $\epsilon$ is the ratio of a typical wavelength $\lambda$ of gravitational waves
and the scale $L$ on which the background curvature changes significantly. Isaacson's high-frequency
approximation thus arises when ${\lambda\ll L}$, i.e. ${\epsilon\ll 1}$. Consequently,
\begin{equation}\label{x1.4}
\begin{array}{rclcrcl}
\gamma_{\mu\nu}&=&O(1)\ ,&&h_{\mu\nu}&=&O(\epsilon)\ ,\\
\gamma_{\mu\nu ,\alpha}&=&O(1)\ ,&&h_{\mu\nu ,\alpha}&=&O(1)\ ,\\
\gamma_{\mu\nu ,\alpha\beta}&=&O(1)\ ,&&h_{\mu\nu ,\alpha\beta}&=&O({\epsilon}^{-1})\ ,
\end{array}
\end{equation}
and one may expand the Ricci tensor in powers of perturbations $h$ as
\begin{equation}\label{2.5}
R_{\mu\nu}(g)=R^{(0)}_{\mu\nu}+R^{(1)}_{\mu\nu}+ R^{(2)}_{\mu\nu}
+ \ldots\ ,
\end{equation}
where
\begin{eqnarray}
&& R^{(0)}_{\mu\nu} \equiv  R_{\mu\nu}(\gamma)\ ,\nonumber \\ %\label{2.6}
&& R^{(1)}_{\mu\nu} \equiv  {\textstyle \frac{1}{2}}\gamma^{\rho\tau}
   \left(h_{\tau\mu ;\nu\rho}+h_{\tau\nu ;\mu\rho}
   -h_{\rho\tau ;\mu\nu}-h_{\mu\nu ;\rho\tau}\right)\ ,\label{2.7}\\
&& R^{(2)}_{\mu\nu} \equiv  {\textstyle  \frac{1}{2}}
   \Big[{\textstyle \frac{1}{2}}h^{\rho\tau}{}_{;\nu}h_{\rho\tau ;\mu}+
   h^{\rho\tau} (h_{\tau\rho ;\mu\nu}+h_{\mu\nu
   ;\tau\rho}-h_{\tau\mu;\nu\rho} - h_{\tau\nu ;\mu\rho})   \nonumber\\
&&  \qquad\qquad +  h^{\tau}{}_{\nu}{}^{;\rho}\left( h_{\tau\mu ;\rho}-h_{\rho\mu;\tau}\right)
-\left(h^{\rho\tau}{}_{;\rho}-{\textstyle \frac{1}{2}}h^{;\tau}\right)
   \left(h_{\tau\mu ;\nu}+h_{\tau\nu ;\mu}-h_{\mu\nu ;\tau}
   \right)\Big]\; .\nonumber
\end{eqnarray}
The semicolons denote covariant differentiation with respect to the background metric $\gamma_{\mu\nu}$,
which is also used to raise or lower all indices. Considering the relations (\ref{x1.4}),
the orders of the terms (\ref{2.7}) are
\begin{equation}\label{rexp}
R^{(0)}_{\mu\nu}=O(1)\ ,\quad
R^{(1)}_{\mu\nu}=O({\epsilon^{-1}})\ ,\quad
R^{(2)}_{\mu\nu}=O(1)\ .
\end{equation}
In the Isaacson high-frequency approximation ${\epsilon \ll 1}$ the dominant
term is $R^{(1)}_{\mu\nu}$ yielding the wave equation
\begin{equation}\label{R1}
\gamma^{\rho\tau}\left(h_{\tau\mu ;\nu\rho}+h_{\tau\nu ;\mu\rho}-h_{\rho\tau ;\mu\nu}-h_{\mu\nu ;\rho\tau}\right)=0
\end{equation}
for the perturbations $h_{\mu\nu}$ on the curved background $\gamma_{\mu\nu}$
(considering the vacuum complete metric $g_{\mu\nu}$ or the case of a  stress-energy tensor which does not contain derivatives of the metric, 
as shown in \cite{podolsky-svitek}).
The two terms of the next order $O(1)$, namely  $R^{(0)}_{\mu\nu}$ and $R^{(2)}_{\mu\nu}$, can be used
to provide the equation for the background (non-vacuum) metric \cite{isaac,podolsky-svitek} which involves
the essential influence of the high-frequency gravitational waves on the background. 
Of course, to obtain a consistent solution, one has to solve both the wave equation and the equation for the background simultaneously. We analyzed this problem explicitly in \cite{podolsky-svitek} for backgrounds with privileged  non-twisting null directions, after simplifying the equations by the WKB approximation.

In this article, however, we wish to concentrate on the principal wave equation (\ref{R1}). Our aim is to obtain evolution of high-frequency gravitational radiation propagating in an arbitrary direction in various  cosmological models. If we impose the TT gauge conditions
\begin{eqnarray}
{h_{\mu\nu}}^{;\nu}&=&0\ ,\label{x1.21}\\
h^\mu_{\ \mu} &=&0\ , \label{x1.22}
\end{eqnarray}
the equation (\ref{R1}) reduces to the following wave equation
\begin{equation}\label{x1.26.1}
{{h_{\mu\nu}}^{;\beta}}_{;\beta}-
2R^{(0)}_{\sigma\nu\mu\beta}\,{h}^{\beta\sigma}-R^{(0)}_{\mu\sigma}\,
{h^{\sigma}}_{\nu}-R^{(0)}_{\nu\sigma}\,{h^{\sigma}}_{\mu}=0\ .
\end{equation}
Consistency of the gauge conditions with the wave equation (\ref{x1.26.1}) was demonstrated in \cite{isaac}.
The conditions (\ref{x1.21}), (\ref{x1.22}) still do not completely exhaust the gauge freedom, and we can thus prescribe the following additional condition
\begin{equation}\label{xx1.30a}
h_{\mu 0}=0
\end{equation}
to further simplify calculations.

\section{High-frequency waves in cosmological models}
Our aim is to investigate solutions of the wave equation (\ref{x1.26.1}), subject to the gauge conditions (\ref{x1.21}) and (\ref{x1.22}), for some relevant cosmological models, namely the Friedmann--Robertson--Walker (FRW), (anti--)de Sitter, and anisotropic Kasner universes.

\subsection{Waves in the FRW models}
A general FRW metric can be written in the standard form
\begin{equation}\label{Kfrw-metric}
\d s^2=a^{2}(\eta)(-\d\eta^2+{}^{3}\gamma_{ij}\,\d x^{i}\d x^{j})\ ,
\end{equation}
where $\eta$ is the conformal time and  ${}^{3}\gamma_{ij}$ is the metric of homogeneous and isotropic three-space with the uniform spatial curvature $K$. Usual choice of the spatial coordinates is
\begin{equation}\label{Kfrw-metric3}
{}^{3}\gamma_{ij}\,\d x^{i}\d x^{j}=\d\chi^2+g^2(\chi)(\d\theta^2+\sin^2\theta\,\d\phi^2)\ ,
\end{equation}
where ${g=\sin\chi}$ for ${K=1}$, ${g=\sinh\chi}$ for ${K=-1}$ and ${g=\chi}$ for ${K=0}$. Let $\gamma_{\alpha\beta}$ and $;$ denote the FRW metric 
(\ref{Kfrw-metric}) and the corresponding covariant derivative, respectively. The covariant derivative with respect to 
the metric ${}^{3}\gamma_{ij}$ will be denoted by $|\,$.

We consider a specific form of the metric perturbations (which has been already used in literature, see e.g. \cite{bardeen}): the time and spatial dependence of the perturbations are separated in the following way,
\begin{equation}\label{frw-perturbation}
h_{\mu\nu}=f(\eta)\, Q_{\mu\nu}\ ,
\end{equation}
where $Q_{\mu\nu}$ satisfies $Q_{\mu 0}=0$, in accordance with the gauge condition (\ref{xx1.30a}). The spatial components of $Q_{ij}$ form a traceless divergenceless tensor (thus guaranteeing that $h_{\mu\nu}$ satisfies the gauge conditions (\ref{x1.21}), (\ref{x1.22})) which is a solution of
\begin{equation}\label{helmholtz}
Q_{ij|l}{}^{|l}+k^2\, Q_{ij}=0\ .
\end{equation} 
Such functions $Q_{ij}$ are called tensor harmonics, and equation (\ref{helmholtz}) is a generalized Helmholtz equation with $k$ representing the wave number which sets the scale of the perturbations relative to the background coordinates. The expansion of perturbations into the tensor harmonics was investigated from the mathematical point of view, e.g., in \cite{d'eath}. 

To simplify the form of the curvature terms in the wave equation (\ref{x1.26.1}) one can use the well-known decomposition of the Riemann tensor \cite{Stephanietal:book}
\begin{equation}\label{Riemann-decomp}
R_{\sigma\mu\nu\beta}=C_{\sigma\mu\nu\beta}+\gamma_{\sigma[\nu}R_{\beta]\mu}-\gamma_{\mu[\nu}R_{\beta]\sigma}
-\frac{1}{3}\gamma_{\sigma[\nu}\gamma_{\beta]\mu}R\ , 
\end{equation}
where $C_{\sigma\mu\nu\beta}$ is the Weyl tensor. For the FRW metric (\ref{Kfrw-metric}) the Weyl tensor vanishes, and one obtains $$R_{ij}=a^{-4}(a\ddot{a}+\dot{a}^{2}+2Ka^{2})\gamma_{ij}\ ,\qquad  R=6a^{-3}(\ddot{a}+Ka)\ ,$$
where $R=\gamma^{\mu\nu}R_{\mu\nu}$. From (\ref{Riemann-decomp}), (\ref{x1.22}),  (\ref{xx1.30a}), and the fact that the spatial 
part $R_{ij}$ of the Ricci tensor is a multiple of $\gamma_{ij}$, we thus derive
\begin{equation}
R_{\sigma\mu\nu\beta}h^{\beta\sigma}=\frac{1}{2}\Big(R^{\nu}{}_{\nu}+R^{\mu}{}_{\mu}-\frac{1}{3}R\Big)h_{\mu\nu}\ 
\end{equation}
(no summation over $\mu,\, \nu$ here). Using the form (\ref{frw-perturbation}) of the perturbation tensor $h_{\mu\nu}$ we derive the following 
expression for the curvature terms in the wave equation (\ref{x1.26.1}),
\begin{equation}\label{curvature-term}
2R_{\sigma\mu\nu\beta}h^{\beta\sigma}+R_{\mu\sigma}h^{\sigma}{}_{\nu}
+R_{\nu\sigma}h^{\sigma}{}_{\mu}=(2a^{-3}\ddot{a}+4a^{-4}\dot{a}^{2}+6Ka^{-2})fQ_{\mu\nu}\ .
\end{equation}
For the covariant d'Alembertian of the perturbation tensor we can write
\begin{equation}\label{frw-dalembert}
h_{\mu\nu}{}^{;\beta}{}_{;\beta}=f^{;\beta}{}_{;\beta}\,Q_{\mu\nu}+2\dot{f}Q_{\mu\nu;0}\gamma^{00}+f(Q_{\mu\nu;00}\gamma^{00}+Q_{\mu\nu;ii}\gamma^{ii})\ ,
\end{equation}
in which 
\begin{eqnarray}
f^{;\beta}{}_{;\beta}=-a^{-2}\ddot{f}-2a^{-3}\dot{a}\dot{f}&,& Q_{\mu\nu;0}=-2a^{-1}\dot{a}Q_{\mu\nu}\ ,\nonumber\\ Q_{\mu\nu;00}=(6a^{-2}\dot{a}^{2}-2a^{-1}\ddot{a})Q_{\mu\nu}&,& Q_{\mu\nu;ii}=Q_{\mu\nu|ii}+2a^{-4}\dot{a}^{2}\gamma_{ii}Q_{\mu\nu}\ .\nonumber
\end{eqnarray} 
Combining equations (\ref{curvature-term}), (\ref{frw-dalembert}), and using the Helmholtz equation (\ref{helmholtz}), the wave equation (\ref{x1.26.1}) is then reduced to the second-order ordinary differential equation for the amplitude of the perturbations $f(\eta)$ depending on the conformal time,
\begin{equation}\label{frw-equation}
\left[\left(\frac{\d^{2}}{\d \eta^{2}}+k^{2}\right)-2\frac{\dot{a}}{a}\,\frac{\d}{\d \eta}+4\frac{\dot{a}^{2}}{a^{2}}+6K\right]f=0\ .
\end{equation}

\subsubsection{Waves in the de~Sitter universe}
As an example we will now solve the equation (\ref{frw-equation}) for the de~Sitter spacetime which is a FRW universe with the cosmological constant
${\Lambda>0}$. Due to its unique symmetry the de~Sitter spacetime admits various specific spacelike sections which provide all three cases of the FRW models (\ref{Kfrw-metric}), (\ref{Kfrw-metric3}) of constant spatial curvature ${K=0,+1}$ or $-1$, see e.g. \cite{eriksen}. 

\begin{figure}[ht]
\begin{center}
   \includegraphics[height=42mm]{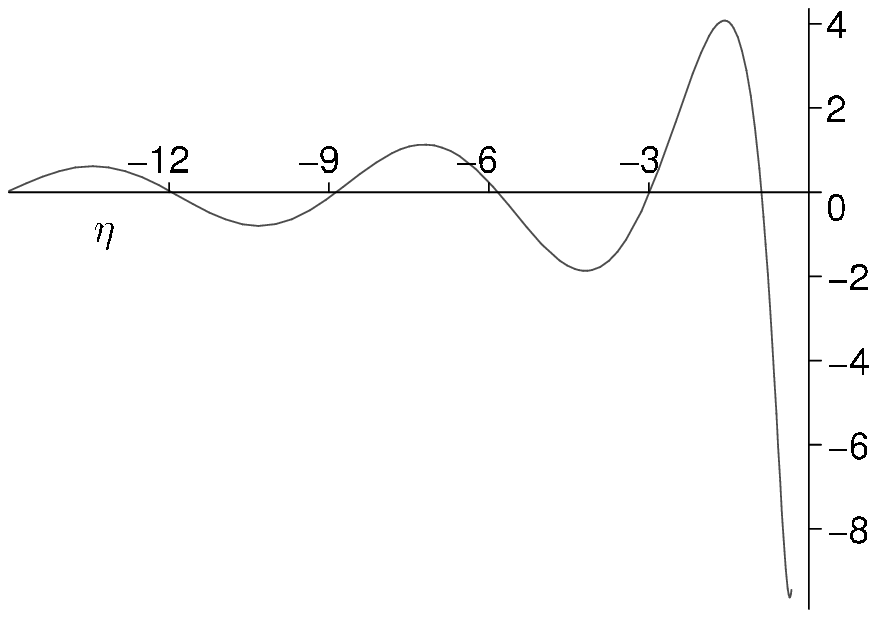}
   \includegraphics[height=42mm]{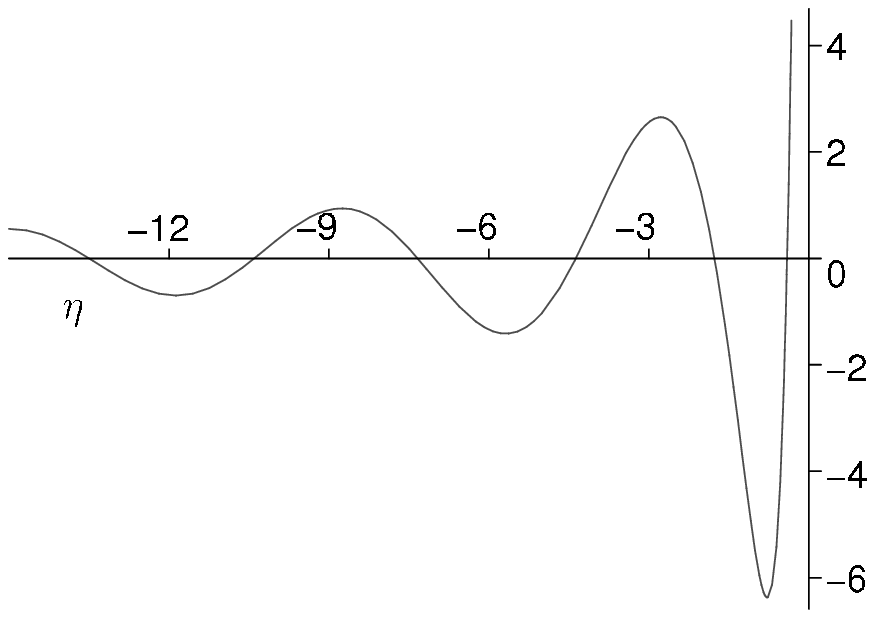}   
\vspace{-3mm}
\hbox{\hspace{0cm} (a) \hspace{5.5cm} (b)} 
\vspace{-5mm}
\end{center}
\caption{\label{fig1}%
The figure (a) is a plot of the function $Re\Big\{\frac{1}{\sqrt{|\eta|}}\,{\rm J}(i\frac{\sqrt{15}}{2},\eta)\Big\}$, and the figure (b) is a plot of $Re\Big\{\frac{1}{\sqrt{|\eta|}}\,{\rm Y}(i\frac{\sqrt{15}}{2},\eta)\Big\}$ representing  evolution of high-frequency gravitational waves in flat de~Sitter universe.}
\end{figure}

The spatially flat $K=0$ form of de Sitter metric is characterized by the expansion function
\begin{equation}\label{deS-K=0}
a(\eta)=-\frac{\alpha}{\eta}\ ,
\end{equation}  
where $\alpha=\sqrt{3/\Lambda}$ and we assume $\eta<0$. The minus sign in (\ref{deS-K=0}) guarantees that $a(t)=a_0\, e^{t/\alpha}$, where $t$~is the comoving synchronous time. This is the standard ``inflationary'' behaviour with the asymptotics $\eta\to 0$ as $t\to\infty$. Inserting the particular form (\ref{deS-K=0}) of the function $a(\eta)$ into equation (\ref{frw-equation}) we arrive at the following solution for the time evolution of the waves
\begin{equation}\label{foursitter}
f(\eta)=\frac{1}{\sqrt{|\eta|}}\left[C_1(k){\rm J}\Big(i\frac{\sqrt{15}}{2},k\eta\Big)+
C_2(k){\rm Y}\Big(i\frac{\sqrt{15}}{2},k\eta\Big)\right]\ ,
\end{equation}
where $C_1, C_2$ are arbitrary functions of $k$, and cylindric Bessel functions of the first kind ${\rm J}(\nu,z)$ and of the second kind ${\rm Y}(\nu,z)$ have an imaginary index ${\nu=i\frac{\sqrt{15}}{2}}$. This expression is in a full agreement with the result obtained previously using synchronous coordinates for the flat de Sitter metric \cite{diplpod}. 
A typical plot, for $k=1$, of the basic modes is given in figure~\ref{fig1}.
Using the relation between the conformal and synchronous times, $|\eta|\sim e^{-t/\alpha}$, we can determine the asymptotic behaviour of the solution as $t\rightarrow\infty$. From the approximate expressions for the Bessel functions, ${\hbox{J}(\nu,z)\sim z^{\nu}}$ and ${\hbox{Y}(\nu,z)\sim z^{-\nu}}$ for ${ z\to 0}$, we infer that with $t\to\infty$ the asymptotic behaviour is periodic, ${\hbox{J}\sim \e^{-i\omega t}}$ and ${\hbox{Y}\sim \e^{i\omega t}}$, where the specific frequency $\omega\equiv\sqrt{\frac{5}{4}\Lambda}$ is the \emph{same} for all the modes $k$, depending on the cosmological constant $\Lambda$ only. The overall amplitude ``envelope'' of the waves (\ref{foursitter}) grows as $\e^{t/2\alpha}$ which should be compared to the growth of the background metric functions $\e^{2t/\alpha}$. In accordance with the cosmic no-hair conjecture the high-frequency waves are thus relatively damped. 
%\vspace{3mm}

\begin{figure}[ht]
\begin{center}
   \includegraphics[height=42mm]{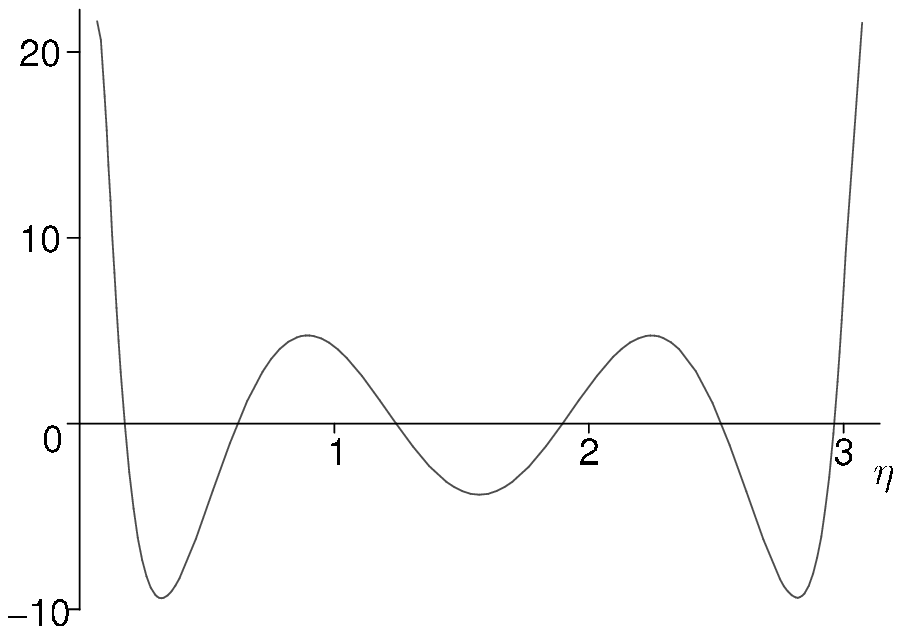} 
   \includegraphics[height=42mm]{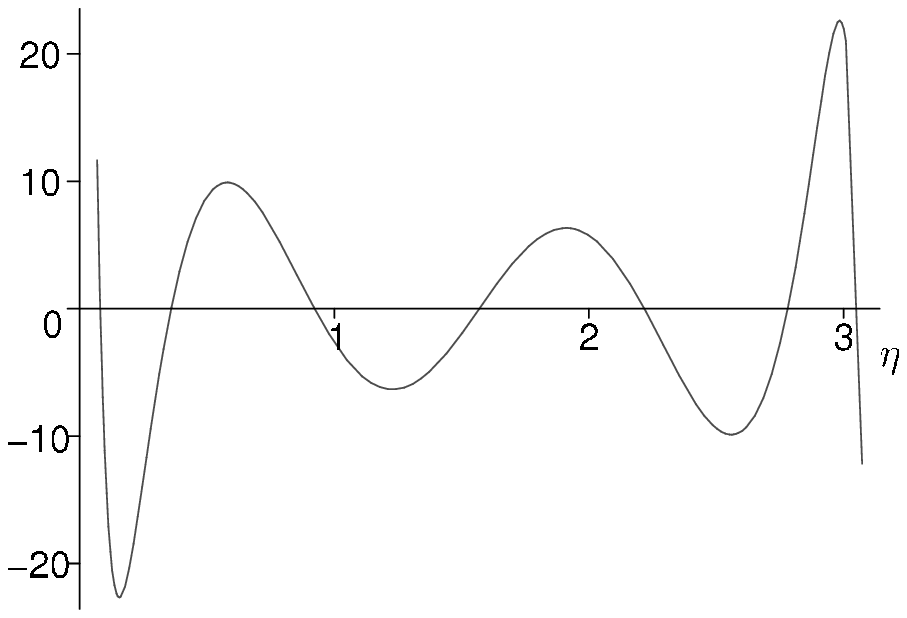}   
\vspace{-3mm}
\hbox{\hspace{0cm} (a) \hspace{5.5cm} (b)} 
\vspace{-5mm}
\end{center}
\caption{\label{fig2}%
The figure (a) is a plot of the function $Re\Big\{\frac{1}{\sqrt{\sin\eta}}\,\hbox{P}(\sqrt{19}-\frac{1}{2},i\frac{\sqrt{15}}{2};\cos\eta)\Big\}$, 
and the figure (b) is a plot of similar expression in which $\hbox{P}$ is replaced by $\hbox{Q}$.}
\end{figure}

The value $K=1$ for metric (\ref{Kfrw-metric}), (\ref{Kfrw-metric3}) corresponds to closed spatial sections  which are three-spheres~$S^{3}$. The expansion function then takes the form
\begin{equation}\label{deS-K=1}
a(\eta)=\frac{\alpha}{\sin\eta}\ ,
\end{equation} 
and the coordinates cover the whole de~Sitter manifold. Solution to the equation (\ref{frw-equation}) with the expansion function (\ref{deS-K=1}) is
\begin{eqnarray}\label{deSsol-K=1}
&&f(\eta)=\frac{1}{\sqrt{\sin\eta}}
 \bigg[C_1(k)\,\hbox{P}\Big(\sqrt{3+k^2}-\frac{1}{2},i\frac{\sqrt{15}}{2};\cos\eta\Big)  \\
&&\hskip21mm+C_{2}(k)\, \hbox{Q}\Big(\sqrt{3+k^2}-\frac{1}{2},i\frac{\sqrt{15}}{2};\cos\eta\Big)\bigg]\ ,\nonumber
\end{eqnarray}
where $\hbox{P}(\nu,\mu;z)$ and $\hbox{Q}(\nu,\mu;z)$ are associated Legendre functions of the first and of the second kind, respectively, which satisfy the differential equation 
\begin{equation}
(1-z^{2})\,y''-2z\,y'+\Big(v(v+1)-\frac{u^{2}}{1-z^{2}}\Big)y=0\ ,
\end{equation}
for $y(z)$. The points $z=1, -1, \infty$ are singularities of this equation (except in some special cases) and ordinary branch points of the Legendre functions in the complex domain. When we take the branch cuts to be $(-\infty,-1)$ and $(1,\infty)$, and if we compose the Legendre functions with the cosine function, as in (\ref{deSsol-K=1}), the singularities occur for ${\eta=0, \pi}$. The sample plots of the basic modes of the solution (for $k=4$) is presented on figure \ref{fig2}. Using the transformation between the conformal time and the synchronous time ${\eta=2\arctan(\e^{t/\alpha})}$ we may determine the asymptotic behaviour of the solution as $t\rightarrow\infty$ which corresponds to $\eta\to\pi$.
From the approximations of associated Legendre functions as $z\to-1$, namely
$\hbox{P}(\nu,\mu;z)\sim A(\nu,\mu)(1+z)^{\mu/2}+B(\nu,\mu)(1+z)^{-\mu/2}$ and 
$\hbox{Q}(\nu,\mu;z)\sim C(\nu,\mu)(1+z)^{\mu/2}+D(\nu,\mu)(1+z)^{-\mu/2}$,
we obtain that the asymptotic behaviour  of both $\hbox{P}$ and $\hbox{Q}$ is periodic in the synchronous time $t$:  it is of the form $\e^{-i\omega t}$ and $\e^{i\omega t}$ as $t\rightarrow\infty$, where the unique frequency $\omega\equiv\sqrt{\frac{5}{4}\Lambda}$ is again common to all the modes. The amplitude of the waves (\ref{deSsol-K=1}) grows as $\sqrt{\cosh(t/\alpha)}$ whereas the de~Sitter metric functions grow as $\cosh^{2}(t/\alpha)$ so that the waves are asymptotically damped with respect to the background.
%\vspace{3mm}

\begin{figure}[ht]
\begin{center}
   \includegraphics[height=40mm]{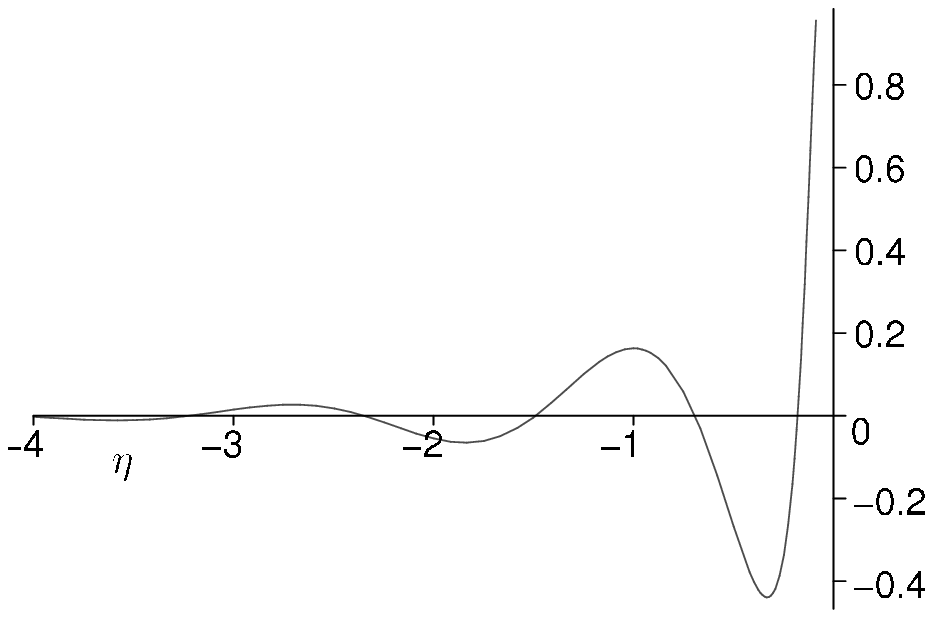}
   \includegraphics[height=40mm]{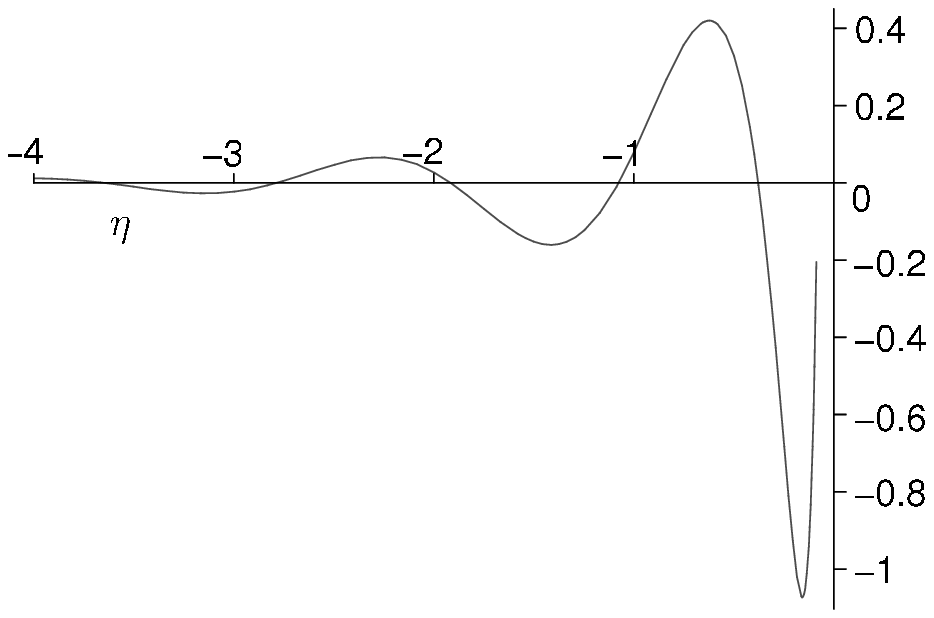}   
\vspace{-3mm}
\hbox{\hspace{0cm} (a) \hspace{5.5cm} (b)} 
\vspace{-5mm}
\end{center}
\caption{\label{fig3}%
The figure (a) is a plot of the function $Re\Big\{\frac{1}{\sqrt{\sinh|\eta|}}\,\hbox{P}(\sqrt{-13}-\frac{1}{2},i\frac{\sqrt{15}}{2};\cosh\eta)\Big\}$, 
and the figure (b) is a plot of an analogous expression, only with $\hbox{P}$ replaced by $\hbox{Q}$.}
\end{figure}

Finally, when $K=-1$ the spatial sections of the de~Sitter spacetime are hyperbolic, and 
\begin{equation}\label{deS-K=-1}
a(\eta)=-\frac{\alpha}{\sinh\eta}\ .
\end{equation} 
Solution to the equation (\ref{frw-equation}) with (\ref{deS-K=-1}) is,
\begin{eqnarray}\label{deSsol-K=-1}
&&f(\eta)=\frac{1}{\sqrt{\sinh|\eta|}}
 \bigg[C_{1}(k)\, \hbox{P}\Big(\sqrt{3-k^{2}}-\frac{1}{2},i\frac{\sqrt{15}}{2};\cosh\eta\Big)\\
&&\hskip23mm +C_{2}(k)\, \hbox{Q}\Big(\sqrt{3-k^{2}}-\frac{1}{2},i\frac{\sqrt{15}}{2};\cosh\eta\Big)\bigg] \ .\nonumber
\end{eqnarray}
Taking the branch cuts again to be $(-\infty,-1)$ and $(1,\infty)$, and composing the Legendre functions with hyperbolic cosine, as in (\ref{deSsol-K=-1}), we obtain the  sample plot (for ${k=4}$) shown in figure \ref{fig3}. Considering the relation to a synchronous time, $\eta=\ln|\tanh(t/2\alpha)|$, we observe that the asymptotic behaviour as $t\rightarrow\infty$ corresponds to $\eta\to0$. If we use the approximation of the Legendre functions 
$\hbox{P}(\nu,\mu;z)\sim E(\nu,\mu)(1-z)^{\mu/2}$, $\hbox{Q}(\nu,\mu;z)\sim F(\nu,\mu)(1-z)^{\mu/2}+G(\nu,\mu)(1-z)^{-\mu/2}$ which is valid near ${z=1}$ we again obtain that the asymptotic behaviour of $\hbox{P}$ and $\hbox{Q}$ in the synchronous time $t$ is asymptotically periodic with the common frequency $\omega=\sqrt{\frac{5}{4}\Lambda}$. The amplitude of the waves grows as $\sqrt{\sinh(t/\alpha)}$ which is slower than the growth of the background metric $\sinh^{2}(t/\alpha)$. 
%\vspace{3mm}
%\newpage

\subsubsection{Waves in the anti--de~Sitter universe}
\begin{figure}[ht]
\begin{center}
   \includegraphics[height=45mm]{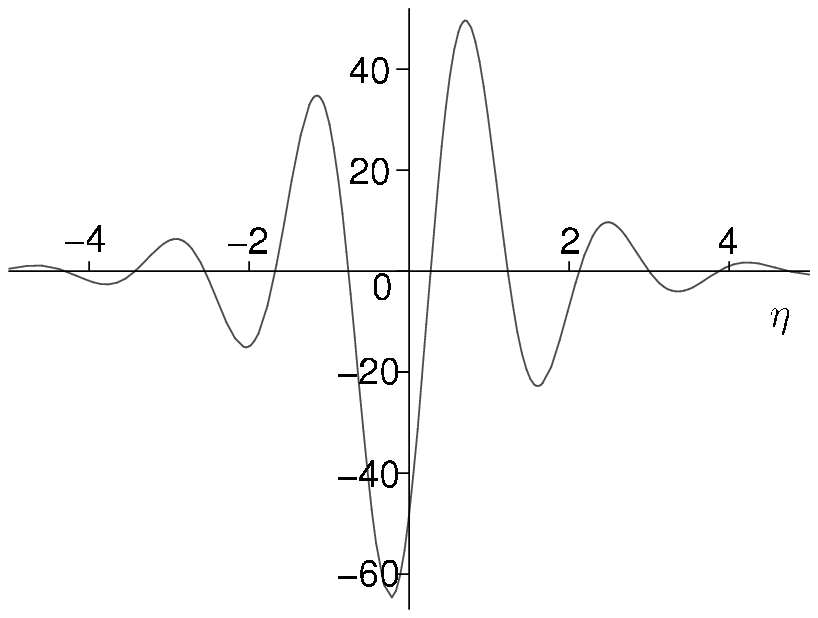}
   \includegraphics[height=45mm]{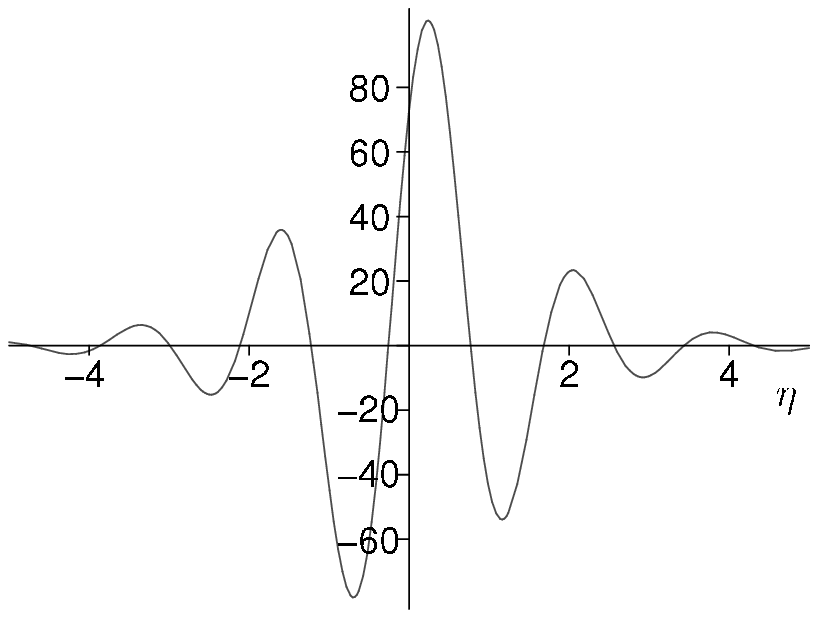}   
\vspace{-3mm}
\hbox{\hspace{0cm} (a) \hspace{5.5cm} (b)} 
\vspace{-5mm}
\end{center}
\caption{\label{fig4}%
The figure (a) is a plot of the function 
$Re\Big\{\frac{1}{\sqrt{\cosh\eta}}\,\hbox{P}(\sqrt{-13}-\frac{1}{2},i\frac{\sqrt{15}}{2};i\sinh\eta)\Big\}$, 
and the figure (b) is a plot of a similar expression, only with $\hbox{P}$ replaced by $\hbox{Q}$. They describe gravitational waves in the anti--de~Sitter universe.}
\end{figure}

Analogously, we obtain the evolution of high-frequency gravitational waves in the anti--de Sitter spacetime represented by the FRW metric with a negative cosmological constant $\Lambda$. The spacetime can be written in the form (\ref{Kfrw-metric}), (\ref{Kfrw-metric3})
with $K=-1$ and the expansion parameter
\begin{equation}\label{AdS-K=-1}
a(\eta)=\frac{\beta}{\cosh\eta}\ ,
\end{equation}
where $\beta=\sqrt{-3/\Lambda}$. However, these coordinates cover only part of the complete manifold. 
Solution to the equation (\ref{frw-equation}) with the expansion function (\ref{AdS-K=-1}) is
\begin{eqnarray}\label{AdSsol-K=-1}
&&f(\eta)=\frac{1}{\sqrt{\cosh\eta}}
 \bigg[C_{1}(k)\, \hbox{P}\Big(\sqrt{3-k^{2}}-\frac{1}{2},i\frac{\sqrt{15}}{2};i\sinh\eta\Big) \\
&&\hskip23mm  +C_{2}(k)\, \hbox{Q}\Big(\sqrt{3-k^{2}}-\frac{1}{2},i\frac{\sqrt{15}}{2};i\sinh\eta\Big)\bigg] \ , \nonumber
\end{eqnarray}
which is the analogue of (\ref{deSsol-K=-1}).
The sample plot of the basic modes (for $k=4$) is presented on  figure \ref{fig4}. The maximum of the envelope is obviously achieved at $\eta=0$, and the waves are damped as $|\eta|\to\infty$.
%\vspace{3mm}

Due to the specific geometry of the anti--de~Sitter spacetime it is not possible to put it into the standard FRW form
(\ref{Kfrw-metric}), (\ref{Kfrw-metric3}) with spatial sections ${K=0}$ and ${K=1}$. To obtain the evolution of possible high-frequency waves in other
standard coordinate forms of the anti--de~Sitter universe, one has to start with the general wave equation (\ref{x1.26.1}). Let us illustrate the procedure in the case of the conformally flat metric
\begin{equation}\label{anti}
ds^{2}=\frac{\beta^{2}}{x^{2}}\left(-\d\eta^{2}+\d x^{2}+\d y^{2}+\d z^{2}\right)\ ,
\end{equation}
(these coordinates cover the whole manifold).
It is easily seen that using the formal complex transformation $\hat{x}=i\eta$, $\hat{\eta}=ix$, $\hat{\alpha}=i\beta$   we obtain the conformally flat metric (\ref{deS-K=0}) of de Sitter spacetime. This offers the possibility to adopt the results obtained for de Sitter spacetime ($K=0$), and to find the high-frequency perturbations for the anti--de Sitter spacetime. Unfortunately, this would imply setting the components $h_{1\mu}$ of the perturbation tensor to zero due to gauge condition $h_{0\mu}=0$ applied in new coordinates. The covariant form of this condition is $h_{\mu\nu}v^{\nu}=0$, where $v^{\nu}$ is the four--velocity of an observer. Therefore, the condition $h_{1\mu}=0$ means that the observer moves faster than the speed of light in the direction of $\frac{\partial}{\partial x}$, in the coordinates of metric (\ref{anti}).

Hence we will attempt to solve the problem directly using the metric (\ref{anti}) and assuming $h_{\mu 0}=0$. The gauge condition $h^{\mu}{}_{\mu}=0$ simplifies to ${h^{i}{}_{i}=0}$ (using the flat space summation convention over the indices of an opposite character).
Non-trivial components of the gauge condition $h_{\mu\nu}{}^{;\nu}$ are ${x\,h^{i}{}_{j,i}-2h_{1j}=0}$.
Applying these conditions on the dynamical wave equations (\ref{x1.26.1}), for perturbations $h_{\mu\nu}$ we obtain
\[h_{11}=0\ ,\quad h_{22}=-h_{33}\ ,\]
\[h_{21,0}=0\ ,\quad h_{31,0}=0\ ,\quad h_{21,2}+h_{31,3}=0\ ,\]
\begin{equation}\label{ads-equations}
x^{2}\,h_{21,k}{}^{k}+4h_{21}=0\ ,\quad
x^{2}\,h_{31,k}{}^{k}+4h_{31}=0\ ,
\end{equation}
\[x^{2}(-h_{22,00}+h_{22,k}{}^{k})+2xh_{22,1}-4xh_{21,2}+4h_{22}=0\ ,\qquad\qquad\]
\[x^{2}(-h_{23,00}+h_{23,k}{}^{k})+2xh_{23,1}-2x(h_{31,2}+h_{21,3})+4h_{23}=0\ .\]
It is obvious that there are only two dynamical degrees of freedom corresponding to $h_{22}=-h_{33}$ and $h_{23}$. The residual components $h_{21}$ and $h_{31}$ are independent of $\eta$ and therefore play the role of (supplementary) boundary conditions. The most natural choice is to put $h_{21}=0=h_{31}$. The solution of the set of equations (\ref{ads-equations}) can thus be interpreted as a purely transversal wave which propagates in the direction of $\frac{\partial}{\partial x}$ and which has two polarizations. By this method we do not obtain waves propagating in a general direction --- this is a consequence of ``anisotropy" of the anti--de Sitter spacetime in the coordinates (\ref{anti}).

With the above choice, the unified form of wave equation for both degrees of freedom $h_{22}=-h_{33}$ and $h_{23}$ is
\[-\frac{\partial^{2} f}{\partial \eta^{2}}+\frac{\partial^{2} f}{\partial x^{2}}+\frac{\partial^{2} f}{\partial
y^{2}}+
\frac{\partial^{2} f}{\partial z^{2}}+\frac{2}{x}\frac{\partial f}{\partial x}+\frac{4}{x^{2}}f=0\ ,\]
where $f$ stands for $h_{22}$ or $h_{23}$. Considering the separation of variables
$f=g(x)\newline \times\exp[{i(-k_{0}\eta+k_{2}y+k_{3}z)}]$ we obtain
\begin{equation}\label{rov-ads}
g''+\frac{2}{x}\,g'+\Big(\frac{4}{x^{2}}+k_{1}^{2}\Big)g=0\ ,
\end{equation}
where $k_{1}^{2}=k_{0}^{2}-k_{2}^{2}-k_{3}^{2}$.
Equation (\ref{rov-ads}) is formally equivalent to (\ref{frw-equation}) for the de Sitter spacetime with $K=0$ (when replacing $x$ with $\eta$, and $k_{1}^{2}$ with $k^{2}$, so that $a=\beta/x$). The general solution is thus a monochromatic high-frequency gravitational wave of the form 
\begin{equation}\label{rov-adsN}
f=\frac{1}{\sqrt{x}}\bigg[C_1(k_{1})\,\hbox{J}\Big(i\frac{\sqrt{15}}{2},k_{1}x\Big)+C_2(k_{1})\,\hbox{Y}\Big(i\frac{\sqrt{15}}{2},k_{1}x\Big)\bigg]
\exp{[i(-k_{0}\eta+k_{2}y+k_{3}z)]}\ .
\end{equation}

Let us finally mention that Isaacson's high-frequency waves in the anti--de Sitter spacetime can also have the form of the Defrise family of exact solutions \cite{defrise,Stephanietal:book}. These metrics were investigated in \cite{podolsky} using the coordinates
\begin{eqnarray}\label{defrise}
&& \d s^{2}=\beta^{2}(\d\theta^{2}+\sinh^{2}{\theta}\d\phi^{2})+8\beta^{2}(\cosh{\theta}+\sinh{\theta}\cos{\phi})^{2}\d u\d v\nonumber\\
&&\qquad\qquad +16\beta^{2}(\cosh{\theta}+\sinh{\theta}\cos{\phi})^{4}d(u)\d u^{2}\ .
\end{eqnarray}
The wavefronts $u=const.$ are two-dimensional hyperbolic surfaces with constant negative curvature $-\beta$ parameterized by $\theta$ and $\phi$.  The anti--de~Sitter background is given by  ${d(u)=0}$, and since the component $\gamma_{uu}$ of the metric is of the form $R(\theta,\phi)d(u)$,
it may also represent high-frequency perturbations for small but rapidly varying function ${d(u)=O(\epsilon)}$. The gauge conditions (\ref{x1.21}), (\ref{x1.22}) are fulfilled identically while the wave equation (\ref{x1.26.1}) is satisfied to the order $O(\epsilon)$.

\subsection{Waves in the anisotropic Kasner universe}
We also investigate Isaacson's high-frequency gravitational waves in cosmological model with lower symmetry, 
namely in the Kasner universe. This is a special case of the Bianchi type~I class of homogeneous but anisotropic spacetimes. Its metric in synchronous coordinates has the standard form \cite{Stephanietal:book}
\begin{equation}
\d s^2=-\d t^2+t^{2p_{1}}\d x^2+t^{2p_{2}}\d y^2+t^{2p_{3}}\d z^2\ ,\label{Kasmet}
\end{equation}
where $p_{1},p_{2},p_{3}$ are constants, which represents a solution of vacuum Einstein's equations if 
\begin{equation}\label{vacuum-kasner}
p_{1}+p_{2}+p_{3}=1\ ,\quad{p_{1}}^{2}+{p_{2}}^{2}+{p_{3}}^{3}=1\ .
\end{equation}
However, we need not assume these relations. It is possible to consider the matter content of the universe described by the energy-momentum tensor that does  not contain a derivative of the metric. This would fulfill the conditions of a generalization of the Isaacson approximation method to non-vacuum spacetimes, as described in \cite{podolsky-svitek}. Let us mention that it has recently been shown \cite{fluid-kasner1,fluid-kasner2} that it is impossible to retain anisotropy when the Kasner universe is filled with a viscous fluid, such that dominant energy condition holds and entropy is non-decreasing. Nevertheless, anisotropy is permitted when it is filled with an ideal fluid satisfying the Zel'dovic equation of state. 

As in the previous calculations we will use $h_{\mu0}=0$ as an additional condition. The traceless gauge condition (\ref{x1.22}) has the form (using the spatial summation convention)
\begin{equation}
t^{-2p_{i}}h^{i}{}_{i}=0\ .
\end{equation}
The gauge condition (\ref{x1.21}) gives the equations
\begin{equation}
p_{i}\, t^{-2p_{i}}h^{i}{}_{i}=0\ ,\ \ t^{-2p_{i}}h^{i}{}_{j,i}=0\ .
\end{equation}
By applying these conditions we simplify the dynamical equations (\ref{x1.26.1}) to \newline
${ p_{i}\, t^{-2p_{i}}h^{i}{}_{j,i}=0}$ and
\begin{equation}\label{waveeq-kasner}
-h_{ij,00}\, t^{2}+\Big(\!-\!\sum_{k}p_{k}+2p_{i}+2p_{j}\Big)t\, h_{ij,0}+t^{2(1-p_{k})}h_{ij,k}{}^{k}-4p_{i}p_{j}\, h_{ij}=0\ 
\end{equation}
(no summation over $i,j$ here).
Using the covariant d'Alembert operator, the differential equation (\ref{waveeq-kasner}) can be rewritten as
$\Box \, h_{ij}+2(p_{i}+p_{j})\,t^{-1} h_{ij,0}-4p_{i}p_{j}\,t^{-2} h_{ij}=0$.
Let us denote an arbitrary component $h_{ij}$ of the perturbation tensor simply as $f$ (even though the wave equation (\ref{waveeq-kasner}) is different for different indices $i,j$) and let us define 
\begin{equation}
A=2(p_{i}+p_{j})-\sum_{k}p_{k}\ ,\quad \ B=p_{i}p_{j}\ .
\end{equation}
We will look for the solutions of (\ref{waveeq-kasner}) in the following special form,
\begin{equation}\label{separation}
f(t,x,y,z)=X(t,x)+Y(t,y)+Z(t,z)\ .
\end{equation}    
Inserting (\ref{separation}) into the wave equation (\ref{waveeq-kasner}), its left-hand side splits into three parts, each depending only on one spatial coordinate. The simplest possible solution is to equate each of these parts to zero, satisfying thus the equation. For example, in the spatial coordinate $x$ we obtain
\begin{equation}\label{Xeq}
\frac{\partial^2 X}{\partial t^2}-\frac{A}{t}\frac{\partial X}{\partial t}-t^{-2p_{1}}\frac{\partial^2 X}{\partial x^2}+\frac{4B}{t^2}\, X=0\ .
\end{equation}
Applying now the one-dimensional Fourier transform in the coordinate $x$ on the equation (\ref{Xeq}) we arrive at the ordinary differential equation
\begin{equation}\label{fourierX}
\frac{\d^2 \tilde{X}}{\d t^2}-\frac{A}{t}\frac{\d \tilde{X}}{\d t}+t^{-2p_{1}}k_{1}^2\, \tilde{X}+\frac{4B}{t^2}\, \tilde{X}=0\ ,
\end{equation}
where $\tilde{X}=\mathcal{F}[X]$. Making an ansatz $\tilde{X}=t^{\frac{1}{2}(A+1)}F(t)$ and using the coordinate transformation $s=\frac{1}{1-p_{1}}k_{1}\, t^{1-p_{1}}$ we obtain the standard form of the Bessel equation
\begin{equation}\label{besseleq}
s^2\, \ddot{G}+s\, \dot{G}+\left(\frac{16B-(A+1)^2}{4(1-p_{1})^2}+s^2\right)\, G=0\ ,
\end{equation}
where ${G(s)=F(t)}$, and the dot denotes differentiation with respect to $s$. Solving (\ref{besseleq}) and transforming this back to $\tilde{X}$ and $t$, the solution of evolution equation (\ref{fourierX}) takes the following form
\begin{eqnarray}
&& \tilde{X}=t^{\frac{1}{2}(A+1)}
 \bigg[C_{1}^{+}(k_{1})\, \hbox{J}\Big({\textstyle\frac{\sqrt{(A+1)^2-16B}}{2(1-p_{1})}},\frac{k_{1}t^{(1-p_{1})}}{1-p_{1}}\Big)   \nonumber\\
&&\hskip18mm
+C_{1}^{-}(k_{1})\, \hbox{Y}\Big({\textstyle\frac{\sqrt{(A+1)^2-16B}}{2(1-p_{1})}},\frac{k_{1}t^{(1-p_{1})}}{1-p_{1}}\Big)\bigg]\ . \label{Xsol}
\end{eqnarray}
For \emph{vacuum} Kasner universe (for which the relations (\ref{vacuum-kasner}) hold) we obtain $(A+1)^2-16B=4(p_{i}-p_{j})^{2}$, and thus the index of the Bessel functions is a real number. Generally, assuming that all $p_{i}$ are positive, it turns out that for $\sum_{k}p_{k}<1$ the index is always real, but for $\sum_{k}p_{k}>1$ it may be imaginary.

Proceeding in the same way for the functions $Y$ and $Z$, the form of the solution (\ref{Xsol}) is reproduced except for the replacement of $k_{1},p_{1},C_{1}^{+},C_{1}^{-}$ with $k_{2},p_{2},C_{2}^{+},C_{2}^{-}\,$, and $k_{3},p_{3},C_{3}^{+},C_{3}^{-}$, respectively. The complete solution to (\ref{waveeq-kasner}) may thus be composed in the following way
\begin{equation}\label{fourier3}
f(t,x,y,z)=\mathcal{F}^{-1}_{3}\left[\tilde{X}(t,k_{1})\, \delta(k_{2})\delta(k_{3})+\tilde{Y}(t,k_{2})\, \delta(k_{1})\delta(k_{3})+\tilde{Z}(t,k_{3})\, \delta(k_{1})\delta(k_{2})\right],
\end{equation}
where $\mathcal{F}^{-1}_{3}$ denotes the inverse Fourier transform in three dimensions, and $\delta$ denotes the Dirac delta function. The high-frequency gravitational waves are then determined by the six functions $C^{\pm}_{j}(k_{j})$, ${j=1,2,3}$. The monochromatic wave with the wave-vector $(k^{0}_{1},k^{0}_{2},k^{0}_{3})$ is obtained by setting ${C^{\pm}_{j}(k_{j})=c^{\pm}_{j}\delta(k_{j}-k^{0}_{j})}$, and has the form
\begin{eqnarray}
&& f(t,x_{1},x_{2},x_{3})=t^{\frac{1}{2}(A+1)}\!\!
   \sum_{j=1,2,3}\bigg[c_{j}^{+}\, 
   \hbox{J}\Big({\textstyle\frac{\sqrt{(A+1)^2-16B}}{2(1-p_{j})}},\frac{k^{0}_{j}t^{(1-p_{j})}}{1-p_{j}}\Big)\label{Kasnercompl}\\
&&\hskip46mm +c_{j}^{-}\, \hbox{Y}\Big({\textstyle\frac{\sqrt{(A+1)^2-16B}}{2(1-p_{j})}},\frac{k^{0}_{j}t^{(1-p_{j})}}{1-p_{j}}\Big)\bigg]\,
\exp{(ik^{0}_{j}x_{j})}\ ,
\nonumber
\end{eqnarray}
where ${x_{1},x_{2},x_{3}}$ stands for ${x,y,z}$.

Since the Bessel function $\hbox{Y}$ is generally divergent when approaching zero, there obviously arises the question of consistency of the above perturbative approach at vicinity of ${t=0}$ when ${p_i<1}$. Asymptotic behaviour of this Bessel function  is 
${\hbox{Y}(\nu,z)\sim z^{-\nu}}$ for ${ z\ll 1}$. Analysis for general Kasner solutions is difficult due to the complicated dependence on three 
parameters $p_{i}$. However, two particular cases can easily be investigated:
\begin{itemize}
\item
For purely imaginary index ${\nu=i\alpha}$ it behaves as a complex unit and therefore it is non-divergent: 
$$\hbox{Y}(\nu,z)\sim e^{-i\alpha\ln z}\ .$$
\item
For vacuum Kasner universe (${\sum p_{i}=1}$) one calculates that
$$\tilde{X} \sim t^{\frac{1}{2}[A+1-\sqrt{(A+1)^2-16B}]}=t^{[(p_{i}+p_{j})-|p_{i}-p_{j}|]}=t^{2p}\ , $$
$$\hbox{where}\quad p=\min\{p_{i},p_{j}\}\ ,$$
which corresponds to the behaviour of the metric coefficients $\sim t^{2p_{j}}$ in (\ref{Kasmet}). The Isaacson approximation is thus
consistent since the amplitudes of the perturbations (i.e., $c_{j}^{+}, c_{j}^{-}$) are small.
\end{itemize} 

\section{Conclusions}
We investigated time evolution of Isaacson's high-frequency gravitational waves in some standard cosmological models. We concentrated mainly on homogeneous and isotropic FRW universes by decomposing the perturbations into tensor harmonics. The Isaacson approach takes the influence of high-frequency waves on the backgrounds into account which is the main difference to other studies of gravitational waves in the above cosmological spacetimes.   

Illustrative explicit results were presented for the maximally symmetric de~Sitter and anti--de~Sitter spacetimes in various forms having different spatial curvatures ${K=0, 1, -1}$. The complete solution representing evolution of gravitational waves which propagate in an arbitrary direction was obtained --- in terms of the Bessel or Legendre functions --- for spacetime metrics with isotropic time-slices.
Interestingly, since the constant parameters $\alpha$ and $\beta$ occur in the conformal factors $a(\eta)$ given by
(\ref{deS-K=0}), (\ref{deS-K=1}), (\ref{deS-K=-1}), (\ref{AdS-K=-1}) only multiplicatively, the corresponding  equation (\ref{frw-equation}) expressed in the conformal time $\eta$ is independent of the cosmological constant in these cases. This explains why the resulting spectra (\ref{foursitter}), (\ref{deSsol-K=1}), (\ref{deSsol-K=-1}), (\ref{AdSsol-K=-1})
do not explicitly depend on the value of $\Lambda$. However, if one introduces the corresponding synchronous time, the dependence on the cosmological constant occurs. For example, in the de~Sitter models the Isaacson wave perturbations \emph{for all modes} $k$ and \emph{for any spatial curvature} $K$ of the background  become periodic asymptotically  as $t\to\infty$ with the common \emph{characteristic frequency} $\omega=\sqrt{\frac{5}{4}\Lambda}$. Moreover, the waves are \emph{exponentially damped} --- relative to the expanding de~Sitter background --- in full accordance with the cosmic no-hair conjecture.

High-frequency gravitational waves in the anisotropic Kasner universe were also studied assuming perturbations with a decoupled spatial dependence. In this case the possible spectrum can be expressed in terms of the Bessel functions. We analyzed the evolution of these perturbations and demonstrated (assuming ${p_i<1}$) that near the initial singularity ${t=0}$ the Isaacson approximation method is consistent. For ${t\to\infty}$  the vacuum perturbations, relative to the background, disappear so that the spacetimes evolve to  homogeneous ones.

\end {document}